\documentclass[aps,prd,12pt,superscriptaddress,nofootinbib,showpacs,eqsecnum]{revtex4-2}

\usepackage{amsmath, amssymb, amscd, mathrsfs}
\usepackage{setspace}
\usepackage{mathtools}
\usepackage{enumerate}
\usepackage{enumitem}
\usepackage{geometry}
\usepackage{graphicx}
\usepackage{xr-hyper}
\usepackage{color, hyperref}
\usepackage{cancel, bbm,wasysym, bm}
\usepackage{upgreek}
\usepackage{empheq}
\usepackage[dvipsnames]{xcolor}
\usepackage{multirow}
\usepackage{easyReview}
\usepackage{pifont}
\usepackage[most]{tcolorbox}
\usepackage{epigraph}
\usepackage[labelfont={bf,footnotesize},textfont=footnotesize]{caption}
\usepackage{soul}
\usepackage{svg}
\usepackage{float}
\usepackage{latexsym}
\usepackage{indentfirst}
\usepackage{fancyhdr}

\allowdisplaybreaks

\newcommand{\toda}[1]{\textcolor{red}{TODO: #1}}
\renewcommand{\toda}[1]{}								

\renewcommand{\(}{\left(}
\renewcommand{\)}{\right)}

\newcommand{\bbb}[1]{\mathbbm{#1}}
\newcommand{\1}{\bbb{1}}
\newcommand{\fc}[1]{\mathcal{#1}}

\newcommand{\ld}{\mathfrak{L}}

\renewcommand{\d}{\partial}

\newcommand{\8}{\infty}

\newcommand{\p}{\hspace*{5ex}}

\newcommand{\umlaut}[1]{\"{#1}}
\newcommand{\schro}{Schr\umlaut{o}dinger }
\newcommand{\<}{\langle}
\renewcommand{\>}{\rangle}

\newcommand{\OO}{\fc{O}}

\newcommand{\twid}[1]{\widetilde{#1}}

\newcommand{\eps}{\varepsilon}

\newcommand{\cmt}[1]{}

\makeatletter
\newcommand{\oset}[3][0ex]{%
	\mathrel{\mathop{#3}\limits^{
			\vbox to#1{\kern-.5\ex@
				\hbox{$\scriptstyle#2$}\vss}}}}
\makeatother

\makeatletter
\let\old@font@info\@font@info
\def\@font@info#1{%
	\expandafter\ifx\csname\detokenize{#1}\endcsname\relax
	\old@font@info{#1}%
	\fi
	\expandafter\xdef\csname\detokenize{#1}\endcsname{}%
}
\makeatother

\numberwithin{equation}{section}

\begin{document}
	
	\title{Emergence of the wavefunction of a non-relativistic quantum particle from QFT}
	\author{Mani L. Bhaumik}
    \affiliation{Department of Physics and Astronomy, University of California, Los Angeles, USA 90095}
	
	\begin{abstract} The nonrelativistic wavefunction of a quantum state that contains all its information is derived directly from the effective quantum fields of the standard model of particle physics, which are the fundamental elements of reality of the universe unveiled to us so far. Consequently, the endless debate about the reality of a wavefunction can now be reasonably put to rest and nonrelativistic quantum mechanics can cogently be considered as a genuine theory.
	\end{abstract}

 	\maketitle
	
	\section{Introduction}
	
	It is quite generally accepted that quantum mechanics is perhaps the most successful theory ever formulated. Experimenters have subjected it to rigorous tests for almost a century and consistently emerged with unfailing triumph. Specifically, the non relativistic quantum theory is involved in a vast array of subjects including physics, chemistry, biology, the emerging quantum technology comprising quantum cryptography, quantum computing, artificial intelligence, and others. 
	
	Yet, despite these spectacular successes, the admired non relativistic quantum mechanics is anchored in, paraphrasing Julian Schwinger in his Nobel Lecture \cite{schwinger} ``dry recital of lifeless axioms'' rather than ``an organic growth and development'' from the ultimate reality revealed to us so far, which is the quantum effective field theory of the standard model of particle physics. In that sense, can it legitimately be called the quantum theory? \cite{wilczek} At the core of non relativistic quantum mechanics is the wave function, which physicists have come to use without fully understanding the physical reality that it describes, leading to contentious debates. This is quite possibly due to the result of how quantum mechanics historically came about. 
	
	The following opinions of the recognized experts provide us with an unshakable trust in the constituents of the fundamental reality. Steven Weinberg \cite{weinberg}, declares that QFT is an unavoidable consequence of the reconciliation of quantum mechanics with special relativity. It has successfully explained almost all experimental observations in particle physics and correctly predicted a wide assortment of phenomena with impeccable precision. By way of many experiments over the years, the QFT of the Standard Model has become recognized as a well-established theory of physics. Weinberg \cite{weinberg2} asserts further that ``The Standard Model provides a remarkably unified view of all types of matter and force (except for gravitation) that we encounter in our laboratories, in a set of equations that can fit on a single sheet of paper. We can be certain that the Standard Model will appear as at least an approximate feature of any better future theory.'' Yet another Physics Nobel Laureate Franck Wilczek underscores on page 96 of \cite{wilczek2}, ``...the standard model is very successful in describing reality -- the reality we find ourselves inhabiting.''
	
	Expressions of such confidence encourage us to anchor our reliance on the QFT of the standard model. One might argue that although the Standard Model accurately describes the phenomena within its domain, it is still incomplete as it does not include gravity, dark matter, dark energy, and other phenomena. However, because of its astonishing success so far, whatever deeper physics may be necessary for its completion would very likely extend its scope without repudiating its current depiction of fundamental reality. 
	
	According to QFT, the fundamental particles which underpin our daily physical reality are only secondary. Each fundamental particle, whether it is a boson or a fermion, originates from its corresponding underlying quantum field. The particles are excitations of quantum fields possessing propagating states of discrete energies, and it is these fields which constitute the primary reality. For example, a photon is a quantum of excitation of the photon field (aka electromagnetic field), an electron is a quantum of the electron quantum field, and a quark is a quantum of a quark quantum field, and so on for all the fundamental particles of the universe.
	
	By far, the most phenomenal step forward made by quantum field theory lies in the stunning prediction that the primary ingredient of everything in this universe is present in each element of spacetime $(t,x,y,z)$ of this immensely vast universe \cite{wilczek2} (page 74). These ingredients are the underlying quantum fields. We also realize that the quantum fields are alive with quantum activity. Furthermore, some of the quantum fluctuations occur at mind-boggling speeds with a typical time period of $10^{-21}$ seconds or less. Despite these wild infinitely dynamic fluctuations, the quantum fields have remained immutable, as required by their Lorentz invariance, essentially since the beginning and throughout the entire visible universe. 
	
	With the detailed preamble presented above, we are ready to proceed now to derive the wave function of a non relativistic quantum particle like the electron from the relativistic quantum field. There are examples galore in physics of similar such procedures. A prominent example is that at large distances and low velocities, Newton's law of gravity can be derived as the appropriate limit of Einstein's General Theory of Relativity. 
	
	Before proceeding further, we need to mention a very important aspect of the various quantum fields. Because of the inherent, spontaneous activity of the quantum fields, any field has a tiny but significant contribution of all the other fields of the standard model. To give a physical depiction of the disturbances of the fields and quantum fluctuations, quoting from Frank Wilczek \cite{wilczek2} (page 89) ``Here the electromagnetic field gets modified by its interaction with a spontaneous fluctuation in the electron field -- or, in other words, by its interaction with a virtual electron-positron pair. [...] The virtual pair is a consequence of spontaneous activity in the electron field. [...] They lead to complicated, small but very specific modifications of the force you would calculate from Maxwell's equations. Those modifications have been observed, precisely, in accurate experiments.'' This is due to the fact that despite the precipitous transitory characteristics of the virtual particles there is an equilibrium distribution \cite{wilczek3}.
	
	This continues on and on, with a ripple in any field disturbing, to a greater or lesser degree, all of the fields with which it directly or even indirectly has an interaction. So, we ascertain that particles are just not simple objects and although we often naively describe them as simple ripples in a single field, that is far from true. Only in a universe with no spontaneous activities -- and no interactions among particles at all -- are particles merely ripples in a single field!
	
	It is particularly important to emphasize again that the quantum fluctuations are transitory but new ones are constantly boiling up to establish an equilibrium distribution so stable that theory contribution to the screening of the bare charge provided the measured charge of the electron to be stable up to nine decimal places \cite{tiesinga} and the electron g-factor results in a measurement accuracy of better than a part in a trillion \cite{xfan}.
	
	Following Dirac, we will now present a detailed calculation of the electron field by a small perturbation of the electromagnetic field. Contributions from the rest of the fields of the standard model will be similar in nature, but smaller.
	
	\section{The Dirac Field and QED}
	
	Weinberg cautions against the use of a wave function of a relativistic particle or system of particles on page 3 of his book on Quantum Field Theory \cite{weinberg}:
	
	\begin{quote}
		It was only later that it became generally clear that relativistic wave mechanics, in the sense of a relativistic quantum theory of a fixed number of particles, is an impossibility. Thus, despite its many successes, relativistic mechanics was ultimately to give way to quantum field theory. 
	\end{quote}
	
	In this paper we shall write down relativistic wave functions, but keep in mind that we effectively deal only with a non-relativistic limit, possibly supplemented with relativistic corrections order by order in $\frac{v}{c}$, where $v$ is the average velocity and $c$ the speed of light. The creation and annihilation of particles is thus suppressed. 
	
	The interacting field theory of Quantum Electrodynamics is described by the Lagrangian density (equation 58.2 of Srednicki \cite{srednicki} or page 78 of \cite{peskin}):
	\begin{gather}
		\ld_{\text{QED}} = \underbrace{-\frac{1}{4}\fc{F}^{\mu\nu}\fc{F}_{\mu\nu}}_{\text{Maxwell}} + \underbrace{i\bar\Psi\gamma^\mu\d_\mu\Psi - m\bar\Psi\Psi}_{\text{Free Dirac}} - \underbrace{e \fc{A}_\mu \bar\Psi\gamma^\mu \Psi}_{\text{Interaction}}. \label{eq:1}
	\end{gather}
	The Maxwell term by itself produces Maxwell's equations in vacuum whereas the free Dirac term by itself produces the Dirac equation for a free particle in vacuum as their equations of motion. The interaction term has the effect of adding a current density $J^\mu = -e \bar\Psi \gamma^\mu \Psi$ to Maxwell's equations and a term $e\gamma^\mu\fc{A}_\mu\Psi$ to the Dirac equation. In our notation:
	\begin{align*}
		&\fc{A}_\mu & &\ \text{is the electromagnetic 4-vector potential}\\
		&\fc{F}_{\mu\nu} = \d_\mu \fc{A}_\nu - \d_\nu \fc{A}_\mu & &\ \text{is the electromagnetic field strength} \\
		&\Psi& &\ \text{is the bispinor Dirac field of spin-1/2 particles}\\
        & & &\ \text{(e.g. electron-positron field)}\\
		&\bar\Psi = \Psi^\dagger\gamma^0& &\ \text{called ``psi-bar'' is the Dirac adjoint of }\Psi \\
		&\d_\mu = \frac{\d}{\d x^\mu} & &\ \text{is the partial derivative with respect to Cartesian } \\
        & & &\ \text{coordinates } x^\mu = (t, x, y, z)^\mu\\
		&\gamma^\mu& &\ \text{are the Dirac matrices} \\
		&m& &\ \text{is the electron mass} \\
		&e& &\ \text{is the magnitude of the electron charge}.
	\end{align*}
	The $\gamma^\mu$ Dirac matrices we take to be given by:
	\begin{gather}
		\gamma^0 = \begin{pmatrix}\1_2  & 0 \\ 0 & -\1_2\end{pmatrix}	\p \p \gamma^a = \begin{pmatrix} 0 & \sigma^a \\ -\sigma^a & 0 \end{pmatrix} 
	\end{gather}
	for $a \in \{1,2,3\}$ where $\1_2$ is the 2x2 identity matrix and $\sigma^a$ are the Pauli-matrices:
	\begin{gather}
		\1_2 = \begin{pmatrix}1 & 0 \\ 0 & 1\end{pmatrix}, \p \sigma^1 = \begin{pmatrix} 0 & 1 \\ 1 & 0 \end{pmatrix}, \p \sigma^2 = \begin{pmatrix} 0 & -i \\ i & 0 \end{pmatrix}, \p \sigma^3 = \begin{pmatrix}1 & 0 \\ 0 & - 1\end{pmatrix}
	\end{gather} 
	Indices are raised and lowered using the Minkowski metric:
	\begin{gather}
		\eta_{\mu\nu} = \text{diag}(1,-1,-1,-1)_{\mu\nu}
	\end{gather}
	and the Einstein summation convention is always used. 

	Equation \eqref{eq:1} is the most general possible relativistic Lagrangian density containing only the electromagnetic field and a massive electrically charged spin-1/2 particle that leads to a renormalizable quantum field theory. The equation of motion for $\Psi$ can be most straightfowardly obtained from \eqref{eq:1} since the Lagrangian contains no $\d_\mu\bar\Psi$. Differentiating \eqref{eq:1}, the equation of motion for $\Psi$ becomes: 
    \begin{gather}
        i\gamma^\mu\d_\mu \Psi - m\Psi = e \gamma^\mu \fc{A}_\mu\Psi. \label{eq:26}
    \end{gather}
    
    The equation of motion for $\fc{A}_\mu$ can be obtained using the Euler-Lagrange equation of motion for the $\fc{A}_\mu$ field:
	\begin{gather}
		\frac{\d\ld}{\d \fc{A}_\mu} = \d_\nu\frac{\d\ld}{\d\d_\nu \fc{A}_\mu}.
	\end{gather}
	In particular, for \eqref{eq:1}:
	\begin{align}
		\frac{\d\ld}{\d \fc{A}_\mu} = -e \bar\Psi \gamma^\mu \Psi, \p \frac{\d\ld}{\d\d_\nu \fc{A}_\mu} = \fc{F}^{\mu\nu}.
	\end{align}
	Together, these result in Maxwell's equations:
	\begin{gather}
		\d_\nu \fc{F}^{\mu\nu} = -e\bar\Psi\gamma^\mu \Psi.
	\end{gather}
	
	\section{The Free Dirac Field}
	
	For $\fc{A}_\mu = 0$, $\Psi$ is free and obeys a linear differential equation which may be solved by Fourier analysis and we write $\Psi = \Psi_0$. Define the function $\twid{\Psi}_0(k)$ as the Fourier transform of $\Psi(x)$:
	\begin{gather}
		\twid{\Psi}_0(k) = \int e^{ik\cdot x} \Psi_0(x) d^4 x
	\end{gather}
	Then, the Fourier transformed Dirac equation becomes:
	\begin{gather}
		(\gamma^\mu k_\mu - m)\twid{\Psi}_0(k) = 0
	\end{gather}
    (equation 3.46 of \cite{peskin}). The most general solution for $\twid{\Psi}_0(k)$ can be found by solving this matrix equation. It is a 4x4 matrix equation. Solutions to this equation only exist for $k^\mu$ so that $k^\mu k_\mu = m^2$. For such $k^\mu$, the equation possesses 4 linearly independent solutions. Those solutions can be characterized by whether they have positive or negative frequency and whether they have positive or negative spin in the direction of some particular axis of choice. With this, the result as seen in equation 3.92 of \cite{peskin}, when returned to $\Psi_0(x)$ gives:
	\begin{align}
		\Psi_0(x) &= \int d^4 k f(k) \sum_s\(u_s(k) b_s(k) e^{-ik\cdot x} + v_s(k)d^\dagger_s(k) e^{ik\cdot x}\), \nonumber \\
		\Psi^\dagger_0(x) &= \int d^4 k f(k) \sum_s\(u^\dagger_s(k) b^\dagger_s(k) e^{ik\cdot x} + v^\dagger_s(k)d_s(k) e^{-ik\cdot x}\) \label{eq:33} 
	\end{align}
	where $s$ is a label that labels the spin-1/2 states, $k$ is the momentum 4-vector, $d^4k$ is the Lorentz-invariant integration measure on momenta, and the Lorentz invariant function $f(k)$ puts the momentum vector on shell. In particular, $f(k)$ is chosen to be:
	\begin{gather}
		f(k) = \frac{\delta(k^2 - m^2)\vartheta(k^0)}{(2\pi)^3}
	\end{gather}
	(where $\vartheta(x)$ is the Heaviside step function) so that when integrating $f(k)$ against any other function $F(k) = F(k^0, \vec k)$:
	\begin{gather}
		\int d^4 k f(k) F(k) = \int \frac{d^3k}{(2\pi)^3 2 k^0} \left.F\(k^0, \vec k\)\right|_{k^0 = \sqrt{m^2+|\vec k|^2}}.
	\end{gather}
	While the right hand side is not manifestly Lorentz invariant, the left hand side is. There are small convention differences between \eqref{eq:33} and equation 3.92 of \cite{peskin} due to our choice to include an extra factor of $\frac{1}{\sqrt{2 k^0}}$ in the definition of the particle creation operators (and to name them as $b, d$ rather than $a, b$) so that they have a simpler transformation property under Lorentz boosts.
	
	The functions $u_s(k)$ and $v_s(k)$ are momentum space wavefunctions for the electron (positive frequency solutions) and the positron (negative frequency solutions) respectively, which are ordinary functions, not operators, and obey the momentum-space Dirac equation (equation 3.46 of \cite{peskin}):
	\begin{gather}
		(\gamma^\mu k_\mu - m)u_s(k) = 0, \p (\gamma^\mu k_\mu + m) v_s(k) = 0.
	\end{gather}
	Importantly, $b_s(k)$ and $d_s(k)$ are the annihilation operators for an electron and positron with momentum $k$ and spin $s$ respectively, while $b^\dagger_s(k)$ and $d^\dagger_s(k)$ are the corresponding creation operators. They obey the anti-commutation relations (equation 3.101 of \cite{peskin}):
	\begin{align}
		&\{b_s(k), b_{s'}(k')\} = \{d_s(k), d_{s'}(k')\} = \{b_s(k), d_{s'}(k')\} = \{b_s(k), d^\dagger_{s'}(k')\} = 0 \nonumber \\
		&\{b^\dagger_s(k), b_{s'}(k')\} = \{d^\dagger_s(k), d_{s'}(k')\} = (2\pi)^3\delta_{s,s'}\delta(k-k').
	\end{align}
	
	The operators for the number of electrons and the number of positrons are separately conserved for the free field case $\fc{A}_\mu = 0$. For interacting QED with $\fc{A}_\mu \neq 0$ only their difference, which is the electric charge operator (equation 3.113 of \cite{peskin}):
	\begin{gather}
		Q = e\int d^4 k f(k) \sum_s(d^\dagger_s(k)d_s(k) - b^\dagger_s(k)b_s(k))
	\end{gather}
	is conserved. The vacuum state $|0\>_\Psi$ for the fermions may be defined by the conditions (equation 3.103 of \cite{peskin}):
	\begin{gather}
		b_s(k)|0\>_\Psi = d_s(k)|0\>_\Psi = 0
	\end{gather}
	for all spin states $s$ and all momenta $k$. The vacuum has zero charge. The Fock space of non-interacting states is the Hilbert space generated by applying strings of products $b^\dagger_s(k)$ and $d^\dagger_s(k)$ operators for different values of $s$ and $k$. For example, a state with three electrons and one positron with momenta $k_i$ and spins $s_s$ is given by $b^\dagger_{s_1}(k_1)b^\dagger_{s_2}(k_2)b^\dagger_{s_3}(k_3)d^\dagger_{s_4}(k_4)|0\>_\Psi$. 
	
	Equation \eqref{eq:33} expresses the general solution to the free Dirac equation as a linear combination of plane-waves. For a spatially compact particle, the coefficients of that linear combination can be selected so that the solution is only nonzero in a finite spatial region. For interacting particles, the solution will take a more complicated form and not simply be a linear combination of plane-waves satisfying the on-shell condition $k^\mu k_\mu = m^2$. 
	
	\section{Free one-particle states and wavefunctions}
	
	The one-particle electron state $|k, s, \text{electron}\>_\Psi$ and positron state $|k,s,\text{positron}\>_\Psi$ are obtained by applying the creation operators to the vacuum (equation 3.106 of \cite{peskin}):
	\begin{gather}
		|k,s,\text{electron}\>_\Psi = b^\dagger_s(k)|0\>_\Psi, \p \p |k,s,\text{positron}\>_\Psi = d^\dagger_s(k)|0\>_\Psi. \label{eq:41}
	\end{gather}
	
	The momentum satisfies the on-shell condition $k^2 = m^2$, and these states have \textit{definite momentum and spin}. The matrix element of the Dirac field between the one-particle states and the vacuum are given by (equation 3.45 of \cite{peskin}):
	\begin{align}
		\psi_{k,s,\text{electron}}(x) &= \<0|\Psi_0(x)|k,s,\text{electron}\>_\Psi = u_s(k) e^{-ik\cdot x} \\
		\psi_{k,s,\text{positron}}(x) &= \<0|\Psi^\dagger_0(x)|k,s,\text{positron}\>_\Psi = v_s^\dagger(k) e^{-ik\cdot x}.
	\end{align}
	The functions $\psi_{k,s,\text{electron}}(x)$ and $\psi_{k,s,\text{positron}}(x)$ are the wavefunctions of these (free) particles. Electrons and positrons are both perfectly legal particles to exist. In fact, for conservation properties like charge and spin, both must be created simultaneously by exciting the electron field with sufficient energy. But both quickly annihilate each other unless some special procedures are employed to separate them as in the LEP collider at CERN. However, our universe has an abundance of electrons (and other matter) compared to positrons (and other anti-matter) because of some lepton-baryong asymmetry in the early universe \cite{burns}. Since all electrons, whether created in the lab today or in the early universe have exactly the same properties, going forward we will focus our attention to electrons, as they are what are relevant to typical experiments.
	
	Instead of a single electron state with fixed momentum $k$ and spin $s$ whose wavefunction is monochromatic and spin polarized, we may construct a wave-packet of one-electron states by linear-superposition of the single electron states (equation 8.1 of \cite{greiner} or described in section 6.2 of \cite{sakurai}):
	\begin{gather}
		|W, \text{electron}\> = \int d^4 k f(k) W(s, k)|k,s,\text{electron}\>_\Psi .
	\end{gather}
	Here $W(s,k)$ is the complex-valued function that describes the wave-packet and the spin mixing. The function $W$ may be taken to be a Gaussian, as visualized in Fig. \ref{fig:1}, but could really have any profile, as considered in \cite{klauber}. By evaluating the matrix element of the Dirac field in this wave-packet state, we obtain the wavefunction for the wave-packet:
	\begin{gather}
		\psi_{W,\text{electron}}(x) = \<0|\Psi_0(x)|W,\text{electron}\>_\Psi = \int d^4 k f(k)\sum_s W(s,k) u_s(k) e^{-ik\cdot x} .
	\end{gather}
	Thus we see that the wavefunctions of the wave-packet state and of the monochromatic states are related by the same wave-packet profile function $W$:
	\begin{gather}
		\psi_{W,\text{electron}}(x) = \int d^4 k\sum_s W(s,k) \psi_{k,s,\text{electron}}(x).
	\end{gather}
	The wavefunction for this wave-packet satisfies the free Dirac equation:
	\begin{gather}
		(i\gamma^\mu \d_\mu - m)\psi_{W,\text{electron}}(x) = 0
	\end{gather}
	for any wave-packet profile $W$. While written here for the case of 3 space dimensions, one can readily truncate to one space dimension (say the $x$ direction) by restricting the momenta $k$ to be in the $x$ direction. One can extract the wavefunction for the non-relativistic limit (plus $v/c$ corrections, and either in $3$ or $1$ space dimensions) by expanding the energy $\hbar k^0$ around $mc^2$ and then expanding the wave function in powers of $v/c$. For consistency one needs to assume that the wave-packet profile $W$ is supported near $\hbar k^0 \sim mc^2$ so as to suppress high energy modes in the wave-packet. To extract the non-relativistic limit and relativistic corrections, one can make use of the Foldy-Woulthuysen transformation \cite{fw}, \cite{costella}. Because we will focus on the non-relativistic limit itself, we simply work directly with components of the Dirac equation.
	
	\section{The electron in a fixed external electromagnetic field}
	
	Next we consider the Dirac field in the presence of an \textit{external} electromagnetic field $\fc{A}_\mu = A_\mu$, where $A_\mu$ is a real-valued function, not an operator. This set-up is suitable for the scattering of electrons in a fixed potential. It is also suitable for electrons bound by a fixed nucleon electric potential, such as in atoms or ionized atoms with few electrons, and for electrons in a fixed magnetic field. For a constant uniform external electric field $E$, Schwinger showed \cite{schwinger2} that the Dirac equation predicts the creation of electron-positron pairs with an amplitude proportional to $\exp(-m^2/(e|E|))$. This function has vanishing Taylor expansion in $e$, so the effect cannot be seen to any order in perturbation theory of small $e$, and the effect is very small for $e|E| \ll m^2$. 
	
	Since we are mostly interested in the non-relativistic limit (possibly extended to include relativistic corrections order by order in the fine structure constant), we shall assume weak external electromagnetic fields $E$ and $B$, more precisely, $e|B|, e|E| \ll m^2$. This assumption is generally valid for scattering problems, but it would not hold for the Coulomb problem whose potential diverges at the location of the electric charge. 
	
	We may then expand the solution $\Psi_A$ to the Dirac equation in the presence of $A_\mu \neq 0$ in terms of the free solution $\Psi_0$. To do so, we return to the Dirac equation \eqref{eq:26} and introduce the free Dirac propagator $S$ (equation 42.13 of \cite{srednicki} up to a conventional factor of $i$):
	\begin{gather}
		(i\gamma^\mu\d_\mu -m)S(x,y) = \1\delta(x-y)
	\end{gather}
	where the derivative on the left acts on the variable $x$. By integrating the Dirac equation against $S(x,y)$, we can recast the differential equation as an integral equation:
	\begin{gather}
		\Psi_A(x) = \Psi_0(x) + e\int d^4 y S(x,y) \gamma^\mu A_\mu(y) \Psi_A(y). \label{eq:52}
	\end{gather}
	
	We then seek a perturbative solution to the Dirac equation in powers of the electromagnetic coupling $e$:
	\begin{gather}
		\Psi_A(x) = \Psi_0(x) + e\Psi_1(x) + e^2\Psi_2(x) + e^3\Psi_3(x) + ... \label{eq:53}
	\end{gather}
	Directly inserting this form into equation \eqref{eq:52} and matching terms at each order in $e$ yields:
	\begin{align}
		\Psi_{(n+1)}(x) = \int d^4 y S(x,y) \gamma^\mu A_\mu(y) \Psi_{(n)}(y).
	\end{align}
	Solving this recurrence relation through order $e^3$ yields:
	\begin{align}
		\Psi_1(x) &= \int d^4 y S(x, y) \gamma^\mu A_\mu(y) \Psi_0(y) \nonumber \\
		\Psi_2(x) &= \int d^4 y d^4 z S(x, y) \gamma^\mu A_\mu(y) S(y, z) \gamma^\nu A_\nu(z) \Psi_0(z) \nonumber \\
		\Psi_3(x) &= \int d^4 y d^4 z d^4 w S(x, y) \gamma^\mu A_\mu(y) S(y, z)\gamma^\nu A_\nu(z) S(z, w) \gamma^\rho A_\rho(w)\Psi_0(w) \label{eq:55}
	\end{align}
	and so on. We can continue to label one-particle electron states $|k,s,\text{electron}\>$ by incoming momentum $k$ and spin $s$ as we did for the free field states in \eqref{eq:41}, and define the wavefunction $\psi_{k,s,\text{electron}}(x)$ to be the matrix element of the field $\Psi_A$ with the vacuum:
	\begin{gather}
		\psi_{k,s,\text{electron}}(x) = \<0|\Psi_A(x)|k,s,\text{electron}\>_\Psi.
	\end{gather}
	Automatically, this wavefunction satisfies the Dirac equation:
	\begin{gather}
		(i\gamma^\mu \d_\mu-m)\psi_{k,s,\text{electron}}(x) = e\gamma^\mu A_\mu(x)\psi_{k,s,\text{electron}}(x). 
	\end{gather}
	Hence we see that the wavefunction is very closely related to the field: namely it is the matrix element between the vacuum and the one-electron state.
	
	Equation \eqref{eq:55} shows how the electron field receives corrections from the electromagnetic field. If we had included other fields that interacted with the electron directly (such as the Higgs boson) or with electromagnetism directly (such as the muon), those fields would make contributions either to the Dirac equation or to the current in Maxwell's equations, which then indirectly correct the electron as well. Thus, in the full standard model of particle physics, all species of fields make some nonzero contribution to the electron's evolution, either directly in its equation of motion or indirectly to the equations of motion of fields that do appear in the electron equations of motion. Because of this, no real particle in the universe consists solely of a free one-particle state of a single species, but is really a complicated mixture of all different fields. 
	
	\section{Quantization of the photon - vacuum fluctuations}
	
	Now we come to the general case, where $\fc{A}_\mu$ is non-zero and generally has an external field part $A_\mu$ and an operator-valued part $\hat A_\mu$ responsible for creating and annihilating photons:
	\begin{gather}
		\fc{A}_\mu(x) = A_\mu(x) + \hat A_\mu(x).
	\end{gather}
	The operator part may be decomposed into creation and annihilation operators $a^\dagger_\sigma(q), a_\sigma(q)$ for photons (which are their own anti-particles) \cite{weinberg3}:
	\begin{gather}
		\hat A_\mu(x) = \int d^4 k f(k) \sum_{\sigma\in\pm 1}\(\eps_{\sigma\mu}(k) a_\sigma(k) e^{-ik\cdot x} + \eps^*_{\sigma\mu}(k) a^\dagger_\sigma(k) e^{ik\cdot x}\)
	\end{gather}
	where $\sigma$ indicates the two polarizations of the photon, and $\eps_{\sigma\mu}(k)$ is the free photon solution to the free Maxwell equations in momentum space for momentum $k$ and polarization $\sigma$ (equation 55.11 of \cite{srednicki}). Due to the U(1) gauge symmetry of electromagnetism, the precise expression for $\eps_{\sigma\mu}(k)$ is gauge dependent, however such gauge dependence drops out of all physical observables (section 57 of \cite{srednicki}). The free photon vacuum state $|0\>_A$ may be defined by:
	\begin{gather}
		a_\sigma(k)|0\>_A = 0
	\end{gather}
	for all $\sigma$ and $k$, while the one-photon state with momentum $k$ and polarization $\sigma$ is given by:
	\begin{gather}
		|k, \sigma,\text{photon}\> = a^\dagger_\sigma(k)|0\>_A.
	\end{gather}
	
	The solution $\Psi_A$, constructed perturbatively in \eqref{eq:53} and \eqref{eq:55} for an external field $A_\mu$, readily generalizes to $\Psi_\fc{A}$ for an operator-valued field $\fc{A}_\mu$ (using the appropriate Feynman propagator $S$, consistent with calculating time-ordered expectation values), and so we have:
	\begin{gather}
		\Psi_{\fc{A}}(x) = \Psi_0(x) + e\Psi_1(x) + e^2\Psi_2(x) + e^3\Psi_3(x) + ... 
	\end{gather}
	where the second, third, and fourth terms are given by:
	\begin{align}
		\Psi_1(x) &= \int d^4 y S(x, y) \gamma^\mu \fc{A}_\mu(y) \Psi_0(y) \nonumber \\
		\Psi_2(x) &= \int d^4 y d^4 z S(x, y) \gamma^\mu \fc{A}_\mu(y) S(y, z) \gamma^\nu \fc{A}_\nu(z) \Psi_0(z) \nonumber \\
		\Psi_3(x) &= \int d^4 y d^4 z d^4 w S(x, y) \gamma^\mu \fc{A}_\mu(y) S(y, z)\gamma^\nu \fc{A}_\nu(z) S(z, w) \gamma^\rho \fc{A}_\rho(w)\Psi_0(w).
	\end{align}
	This expansion is responsible for a whole variety of interesting physical processes, but here we will concentrate on the wavefunction aspects only. The \textit{full} vacuum of the combined photon-electron system is:
	\begin{gather}
		|0\> = |0\>_\Psi\otimes|0\>_A
	\end{gather}
	where the tensor product $\otimes$ is just a mathematical way of expressing the fact that we put the electron and photon Hilbert spaces together. The \textit{full} one-electron state (which contains no photons) may be defined by:
	\begin{gather}
		|k,s,\text{electron}\> = |k,s,\text{electron}\>_\Psi\otimes|0\>_A.
	\end{gather}
	
	\textbullet\ \ First, consider setting the external field to zero ($A_\mu = 0$) retaining only the operator field. We can again define a wave function as the matrix element of the Dirac field between the full vacuum $|0\>$ and the full one electron state:
	\begin{gather}
		\psi_{k,s,\text{electron}}(x) = \<0|\Psi_{\hat A}(x)|k,s,\text{electron}\>.
	\end{gather}
	Since the vacuum expectation value of an odd number of photon creation and annihilation operators vanishes, the only contributions to the order in $e$ we have computed are:
	\begin{gather}
		\psi_{k,s,\text{electron}}(x) = \<0|\Psi_0(x)|k,s,\text{electron}\> + e^2 \<0|\Psi_2(x)|k,s,\text{electron}\> + \OO(e^4).
	\end{gather}
	The first is just the wavefunction for a free electron, while the second contains the vacuum polarization effect due to the exchange of one photon. This gives the \textit{self-energy} contribution computed first by Schwinger and it requires renormalization to be handled properly. 
	
	\textbullet\ \ Second, consider now turning on an external field $A_\mu$ and retaining only its linear effects. Only $\Psi_1$ and $\Psi_3$ now contribute (to linear order in $A$):
	\begin{gather}
		\psi_{k,s,\text{electron}}(x) = e\<0|\Psi_1(x)|k,s,\text{electron}\> + e^3 \<0|\Psi_3(x)|k,s,\text{electron}\> + \OO(e^5).
	\end{gather}
	The contribution $\Psi_1$ simply gives the effect of an external source already computed in \eqref{eq:55}. The really interesting and new contribution comes from $\Psi_3$, specifically from:
	\begin{gather}
		\Psi_3(x) = \int d^4 y d^4 z d^4 w S(x, y) \gamma^\mu \hat A_\mu(y) S(y, z) \gamma^\nu A_\nu(z) S(z, w) \gamma^\rho \hat A_\rho(w) \Psi_0(w).
	\end{gather}
	This gives the \textit{vertex operator renormalization} which was also computed by Schwinger \cite{schwinger3} and gives the value of the anomalous magnetic moment $g-2 = \frac{\alpha}{2\pi}$. 
	
	\section{Effective interaction due to one-loop vacuum fluctuations}
	
	Once the effects of vacuum fluctuations of the electromagnetic field have been included, as we did here by including $\Psi_2$ or $\Psi_3$, then the corresponding wavefunction no longer satisfies a simple Dirac equation. Although the field $\Psi$ still satisfies the Dirac equation:
	\begin{gather}
		(\gamma^\mu(i\d_\mu - e A_\mu - e\hat A_\mu)-m)\Psi = 0
	\end{gather}
	the presence of the operator-valued $\hat A_\mu$ prevents one from writing down a simple equation for the matrix element $\<0|\Psi(x)|k,s,\text{electron}\>$. The modifications to the Dirac equation need to be computed using the methods of quantum field theory and are usually highly non-trivial. For example, the lowest order effect of the term $\Psi_3$ is the anomalous magnetic moment of the electron, whose modification to the Dirac equation is given by:
	\begin{gather}
		\(\gamma^\mu(i\d_\mu - e A_\mu) - m - \frac{g-2}{4m} e\fc{F}_{\mu\nu} S^{\mu\nu}\)\Psi = 0, \p \p S^{\mu\nu} = \frac{i}{4}[\gamma^\mu,\gamma^\nu]
	\end{gather}
	plus higher order terms in $\fc{F}/{m^2}$ which we shall neglect here. The above effective Dirac equation now depends only on the background field $A_\mu$. 
	
	\section{The Dirac wavefunction from the Dirac field}
	
	The background field $A_\mu$ is arbitrary and does not have to satisfy Maxwell's equations with the Dirac fermion as its source. Thus, to describe the solutions to the equation, we shall assume that $A_\mu(x)$ is localized in space and time, and vanishes (or at least tends to zero rapidly enough) as space and time coordinates are taken toward $\8$. In that case, the solutions may be labeled by the momentum $k$ and spin label $s$ of the asymptotic solution as $x^0 \to -\8$:
	\begin{gather}
		\Psi(x) = \int d^4 k f(k)\sum_s\(b_s(k)\psi_{b s}(k, x) + d^\dagger_s(k) \psi_{d s}(k,x)\)
	\end{gather} 
	where $\psi_{bs}(k,x)$ and $\psi_{ds}(k,x)$ both satisfy the modified Dirac equation:
	\begin{gather}
		\(\gamma^\mu(i\d_\mu - e A_\mu) m - \frac{g-2}{4m}e\fc{F}_{\mu\nu} S^{\mu\nu}\)\psi_{bs}(k,x) = 0
	\end{gather}
	and correspondingly for $\psi_{ds}(k,x)$. As $x^0\to-\8$ the Dirac equation becomes approximately non-interacting, thanks to our assumption that the field $A_\mu(x)$ vanishes in this limit, and we identify the incoming solutions with the free solutions $u_s(k)$ and $v_s(k)$ as follows:
	\begin{align}
		\lim_{x^0\to-\8}&\(e^{ik\cdot x} \psi_{bs}(k,x)\) = u_s(k) \nonumber \\
		\lim_{x^0\to-\8}&\(e^{-ik\cdot x}\psi_{ds}(k,x)\) = v_s(k).
	\end{align}
	
	One may now define the vacuum as usual:
	\begin{gather}
		b_s(k)|0\> = d_s(k)|0\> = 0
	\end{gather}
	for all $s$ and all $k$, as well as one-particle states:
	\begin{gather}
		|k,s,\text{electron}\> = b^\dagger_s(k)|0\>, \p \p |k,s,\text{positron}\> = d^\dagger_s(k)|0\>.
	\end{gather}
	By construction, the wave function is given as the matrix element of the field between the vacuum and the corresponding 1-particle state:
	\begin{gather}
		\psi_{bs}(k, x) = \<k,s,\text{electron}|\Psi(x)|0\>.
	\end{gather}
	A wave-packet may be constructed from this mono-chromatic solution just as we did in the free-field case. For arbitrary background electromagnetic field $A_\mu$, the above wavefunctions will not form a complete set of states as for example the creation of electron-positron pairs has not been taken into account, since it involves not just one-particle states, but also three-particle states and so on. To obtain a consistent truncation to one-particle states, one must take the non-relativistic limit, in which all exciting energies are much smaller than the electron rest-energy $mc^2$, and where pair creation is forbidden on purely energetic grounds. This limit is standard material given, for example, in Dirac's book \cite{dirac}. We briefly review it here to leading order. To work out corrections to higher order in $v/c$ one usually appeals to the Foldy-Wouthuysen \cite{fw} transformation, but to the leading order considered here this tool is not needed.
	
	\section{The non-relativistic limit of the Dirac equation}
	
	In the non-relativistic limit, the energy $k^0$ of the electron is only slightly different from its rest mass, $|k^0-m| \ll m$. A consistent approximation requires that, in the proper units, the field $A_\mu$ or more appropriately the field strength $\fc{F}_{\mu\nu}$ is small compared to $m$. Using our conventions for the Dirac matrices, the spin matrices $S^{\mu\nu}$ are then given as follows:
	\begin{gather}
		S^{0a} = \frac{i}{2}\begin{pmatrix}0 & \sigma^a \\ \sigma^a & 0 \end{pmatrix}, \p \p S^{ab} = \frac{1}{2}\epsilon^{abc}\begin{pmatrix}\sigma^c & 0 \\ 0 & \sigma^c\end{pmatrix}
	\end{gather}
	so that:
	\begin{gather}
		\frac{1}{2}\fc{F}_{\mu\nu}S^{\mu\nu} = \begin{pmatrix} \vec B\cdot \vec S & i \vec E\cdot \vec S \\ i\vec E\cdot\vec S & \vec B\cdot\vec S\end{pmatrix}, \p \vec S = \frac{\vec\sigma}{2}, \p \kappa = \frac{g-2}{2m}e
	\end{gather}
	where $\vec E$ and $\vec B$ are the electric and magnetic fields of Maxwell's equations. To obtain the non-relativistic limit, we factor out the leading oscillatory time-dependence of the wavefunction due to the rest mass energy of the electron and decompose the 4-component spinor into two 2-component spinors which diagonalize $\gamma^0$:
	\begin{gather}
		\psi_{bs}(k,x) = e^{-imx^0}\begin{pmatrix}\psi_+ \\ \psi_-\end{pmatrix}.
	\end{gather}
	The decomposition into 2-component spinors is necessary because in the non-relativistic limit, the spin-1/2 electron has only two components rather than the four present in the Dirac equation. Then, the Dirac equation decomposes as follows:
	\begin{align}
		(i\d_t + m -e A_0)\psi_+ &+ (i\d_a - e A_a)\sigma^a\psi_- - m\psi_+ - \kappa B\cdot S\psi_+ - i\kappa E\cdot S\psi_- = 0 \\
		(i\d_t + m -e A_0)\psi_- &+ (i\d_a - e A_a)\sigma^a\psi_+ + m\psi_- + \kappa B\cdot S\psi_- + i\kappa E\cdot S\psi_+ = 0
	\end{align}
	where the contribution $m$ inside the first set of parentheses is the result of differentiating the oscillator prefactor. The masses cancel in the first equation and double up in the second, so that we obtain the simplified forms:
	\begin{align}
		(i\d_t -e A_0)\psi_+ &+ (i\d_a - e A_a)\sigma^a\psi_-- \kappa B\cdot S\psi_+ - i\kappa E\cdot S\psi_- = 0 \\
		(i\d_t + 2m -e A_0)\psi_- &+ (i\d_a - e A_a)\sigma^a\psi_+ + \kappa B\cdot S\psi_- + i\kappa E\cdot S\psi_+ = 0.
	\end{align}
	Next, we use the fact that $\d_0, A_0$ and $\vec B$ are negligible compared to $m$ in order to simplify the second equation, so that in the non-relativistic limit:
	\begin{gather}
		2m\psi_- + (i\d_a - eA_a)\sigma^a\psi_+ + i\kappa \vec E\cdot\vec S\psi_+ = 0.
	\end{gather}
	Specializing to the case of a purely magnetic field simplifies the equations, and using the above equation to eliminate $\psi_-$ in the first equation, we obtain:
	\begin{gather}
		(i\d_t - eA_0)\psi_+ = \frac{1}{2m}(i\d_a - e A_a)\sigma^a (i\d_b - e A_b)\sigma^b \psi_+ + \kappa\vec B\cdot\vec S\psi_+.
	\end{gather}
	Using the relations:
	\begin{gather}
		[(i\d_a - e A_a), (i\d_b - eA_b)] = -ie \fc{F}_{ab}, \p [\sigma^a, \sigma^b] = 2i\epsilon^{abc}\sigma^c
	\end{gather}
	we can further simplify the equation into:
	\begin{gather}
		(i\d_t - e A_0)\psi_+ = \frac{1}{2m}(i\nabla-e\vec A)^2 \psi_+ + \frac{ge}{2m} \vec B\cdot\vec S\psi_+ \label{eq:911}
	\end{gather}
	which is the non-relativistic \schro equation in the presence of a magnetic field, also referred to as the Pauli equation, and incorporating the anomalous magnetic moment of the electron. Therefore, \textit{the non-relativistic single electron wavefunction arises naturally from matrix elements in quantum field theory.} 

	\section{Restriction to One Spatial Dimension}
	
	In one spatial dimension, there is no magnetic field and we can choose a gauge in which $\vec A = 0$. Re-baptizing $\psi_+ \to \psi$ and $eA_0 \to V$, we obtain the familiar one-dimensional \schro equation with potential $V(t,x)$:
	\begin{gather}
		i\hbar\frac{\d\psi}{\d t}(t,x) = -\frac{\hbar^2}{2m}\frac{\d^2\psi}{\d x^2}(t,x) + V(t,x)\psi(t,x)
	\end{gather}
	where we restored the dependence on $\hbar$ and allowed for time-dependence in the potential $V$ resulting from the time-dependence of a general electric potential $A_0$.
	
	To make contact with equation (1.3.2) of Weinberg \cite{weinberg4} and (10-9) of Klauber \cite{klauber}, we will make some further assumptions on the potential $V$ to simplify the problem without losing its physical significance:
	\vspace{2mm}
	\setlist{nolistsep}
	\begin{enumerate}
		\item $V$ is independent of time,
		
		\item as a function of $x$, the potential $V(x)$ tends to zero as $x\to\pm\8$, 
		
		\item the potential is positive everywhere so that there are no bound states in the spectrum. (Including the case with bound states requires a bit of extra math.)  \\
	\end{enumerate}
	
	We now investigate the solution to the problem as follows. For $x\to-\8$, the potential vanishes, and the \schro equation is then solved in the limit $x\to-\8$ by a linear combination of plane-waves $e^{ikx}$ for any real value of $k$. In the absence of bound states, as we assumed is the case, these solutions give the complete spectrum. We shall denote the solution asymptotic to $e^{ikx}$ as $x\to-\8$ by $\psi_k(x)$. The time-dependence of this solution is given by $e^{-i\omega(k)t}$ so that altogether the solution with the incoming momentum $k$ is given by:
	\begin{gather}
		\psi(t,x) = \psi_k(x) e^{-i\omega(k)t}, \p \omega(k) = \frac{\hbar k^2}{2m}.
	\end{gather}
	
\begin{figure}[h!] 
		\begin{center}\includegraphics[scale=.3]{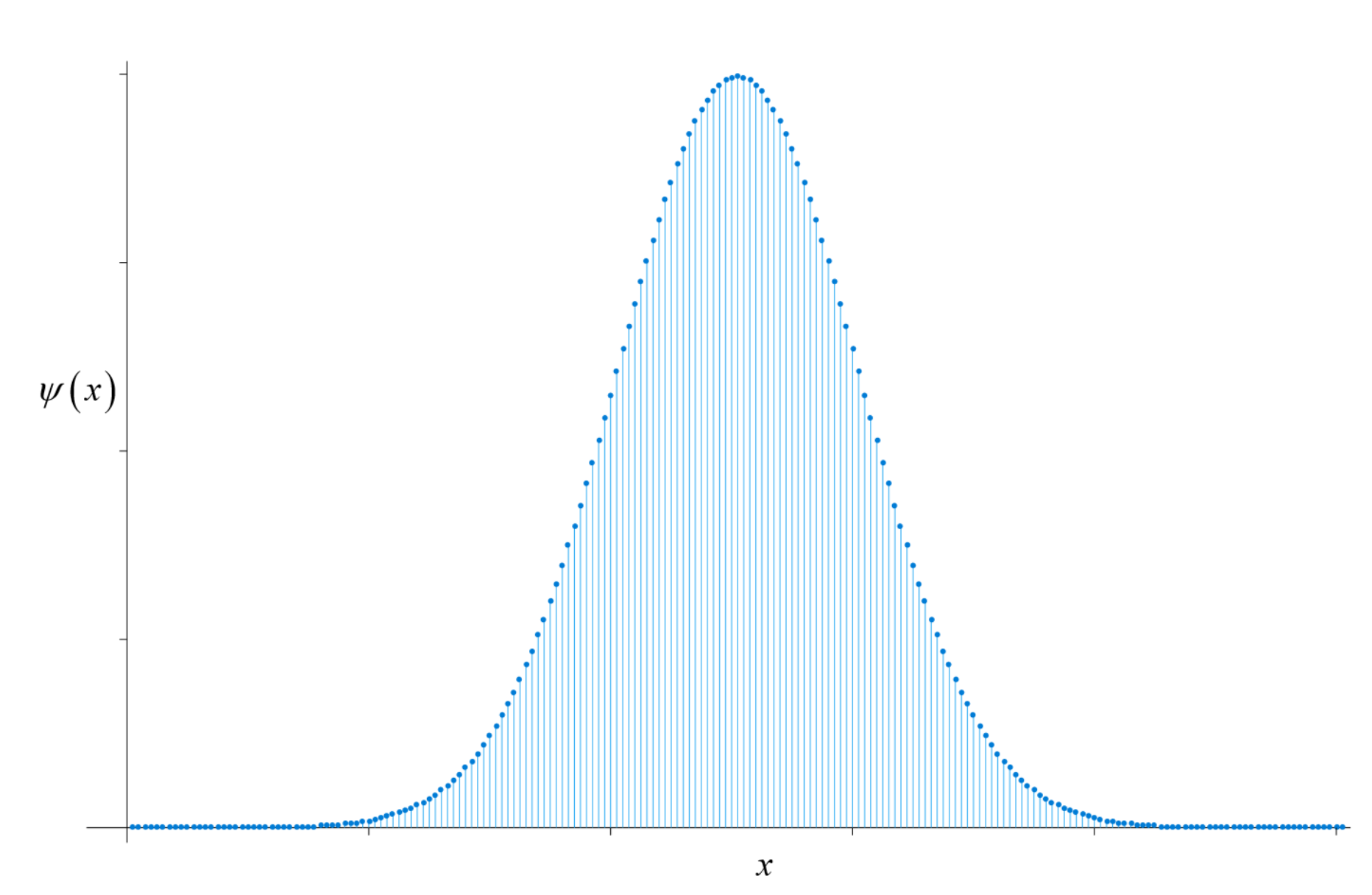}\end{center}
		\caption{Plot of Gaussian wave packet function of an electron portrayed in position space.}
		\label{fig:1}
	\end{figure}

	Any wavefunction can be expressed as a linear combination of these functions $\psi_k(x)$. In particular, we can construct a wave-packet by integrating the above solutions against a wave-packet profile $g(k)$ following Weinberg's notation, which gives the following solution to the \schro equation:
	\begin{gather}
		\psi(t,x) = \int_{-\8}^\8 dk g(k) \psi_k(x) e^{-i\omega(k) t}
	\end{gather}
	where we have $\omega(k)$ exactly given as before. At a particular time, this function $g(k)$ could be chosen so that $\psi(t, x)$ at that time is a Gaussian for example, as in Fig. \ref{fig:1}.
	
	In the limit $x\to-\8$, we have $\psi_k(x) \to e^{ikx}$ and we recover Weinberg's formula (1.3.2) \cite{weinberg4}:
	\begin{gather}
		x\to-\8 \implies \psi(t,x) \to \int_{-\8}^\8 dk g(k) e^{ikx-i\omega(k)t}.
	\end{gather}
	Actually Weinberg gives this expression for the wavefunction of a wave-packet really only in the case of a free particle, where we have exactly $\psi_k(x) = e^{ikx}$ for all $x$. For the interacting case, one needs to use the asymptotic behavior as we have done.
	
	The formula (10-9) in Klauber also gives a wave-packet but no time-dependence is included, and as a result, it should be viewed as an initial condition for the subsequent time-evolution under the \schro equation. The time-dependence of the wave-packet is then governed by the \schro equation, but this does not seem to be made explicit in Klauber. Also, Klauber's analysis is done only for a scalar field, whereas we have performed it for a spinor field. 
	
	\section{Concluding Remarks}
	
	Despite the general impression of some quantum field theorists that the derivation of the wavefunction of nonrelativistic quantum mechanics at best would be very difficult, we have shown here persuasively that it can indeed be achieved. Such a direct derivation leaves little doubt about the reality of the wavefunctions of nonrelativistic quantum mechanics since it emerges genuinely from the elements of the ultimate reality of quantum fields. 

    The development of quantum mechanics is replete with a notable trend. Because of the sheer novelty of the subject so remarkably different from the established classical physics, the pioneers of the development of quantum physics utilized a procedure quite frequently with notable success. Due to a thorough lack of experience with the precepts of the newly emerging subject of quantum mechanics, an empirical model was fashioned first to accommodate the observed information. The empirical model was then amended to accommodate a more realistic version from a deeper understanding gained from subsequent revelations. 
	
	This successful procedure started almost from the beginning with the proposal of a quantum by Max Planck. Out of sheer frustration of not being able to match the characteristics of blackbody radiation to his equation, Planck introduced the indivisible radiation quantum believing it was just a necessary mathematical oddity without having any reality whatsoever. Five years later, Einstein argued for the reality of the quantum from the results of photo-electric effect.
	
	Citing another example, Neils Bohr crafted the first atomic model with discrete electron orbits to fit the observed spectral data. With de Broglie's matter wave hypothesis and its subsequent experimental verification, Erwin \schro demonstrated Bohr's discrete atomic orbits to be the standing wave patterns of matter waves, and the list continues. A similar situation presented itself for the emergence of the wave packet function.
	
	The matter wave proposed by de Broglie was simply a plane wave while a quantum particle like an electron is localized. However, a wave packet function consisting of a Fourier integral with the appropriate weighted linear combinations of plane-waves can be assembled that would represent a localized particle.
	
	Erwin \schro formulated a version of quantum mechanics that was based on such waves. He wrote down a wave equation that governs how the waves depend on space and evolve in time. Even though the equation is correct, the precise interpretation of what the wave meant was still missing. Initially \schro thought incorrectly that the wave represented the charge density of the electron. Using the ad hoc Born rule, Max Born correctly interpreted Schr\umlaut{o}dinger's wave as a probability amplitude. By amplitude we mean that the wave magnitude must be squared to obtain the desired probability.
	
	There are many historically baffling aspects of the wave function. The wave particle duality, uncertainty principle, simultaneous existence of a quantum particle in more than one place, and the measurement problem can be mentioned just to give some examples. Fortunately, many of these conundrums disappear when the wave function of a quantum particle is innately derived from the fundamental reality of the universal quantum fields of the Standard Model of particle physics. 
 
	Based on the assumption of a successful derivation of the reality of the wavefunction, the author has presented a series of papers \cite{bhaumik1}--\cite{bhaumik5} that goes a long way to illustrate that quantum mechanics can be a real theory and not just based on axioms. \\
	
	{\noindent\Large\textbf{Acknowledgments}}
	
	The author would like to express deep gratitude to Eric d'Hoker for his gracious endeavors in helping to formulate this article. Beneficial discussions with Trevor Scheopner and Zvi Bern are much appreciated. Mark Srednicki's valuable comments are earnestly recognized as well. \\

\newpage

\providecommand{\href}[2]{#2}\begingroup\raggedright\endgroup
 
\end{document}